\documentclass[10pt,letterpaper,twocolumn]{article}
\usepackage[english]{babel}
\usepackage{graphicx}

\newcommand{\alt}{\mathbin{\lower 3pt\hbox
   {$\rlap{\raise 5pt\hbox{$\char'074$}}\mathchar"7218$}}}
\newcommand{\agt}{\mathbin{\lower 3pt\hbox
   {$\rlap{\raise 5pt\hbox{$\char'076$}}\mathchar"7218$}}}

\begin{document}

\setcounter{footnote}{0}
\setcounter{equation}{0}
\setcounter{figure}{0}
\setcounter{table}{0}

\title{\large\bf Comment on "Can disorder really enhance
superconductivity?"}

\author{\small  I. M. Suslov \\
\small Kapitza Institute for Physical Problems,
\\  \small Moscow, Russia \\{} \\
\parbox{120mm}{\footnotesize \,The paper by Mayoh and
Garcia-Garcia [arXiv:1412.0029v1] is entitled
"Can disorder really enhance
superconductivity?". In our opinion, the answer given by the
authors is not satisfactory. We present the alternative picture.
 }
}

\date{}

\maketitle

The paper by Mayoh and Garcia-Garcia \cite{1} is entitled "Can
disorder really enhance superconductivity?". In our opinion, the
answer given by the authors is not satisfactory and we present
the alternative picture.
\vspace{2mm}

{\it Mean field solution.}
A basis for description of the spatially inhomogeneous
superconductivity is given by the Gor'kov equation for
the order parameter   $\Delta(r)$
$$
\Delta(r) =  \int  K ( r,  r')
\Delta( r') d^d r'
\eqno(1)
$$
with the kernel  $K(r,r')$ satisfying the sum rule \cite{2}
$$
\int  K (r,r') d^d r' = g \nu_F(r)
\ln\frac{1.14\omega_0}{T} \,,
\eqno(2)
$$
where $g$ is the Cooper interaction constant,  $\nu_F(r)$ is the
local density of states at the Fermi level, $\omega_0$ is a
cut-off frequency, and $d$ is a dimension of space.

The Anderson theorem \cite{3} follows from Eq.1 under assumption
of a self-averaging order parameter, when
$\Delta(r)$ and $K(r,r')$  can be independently averaged
over disorder. Since  $\langle\Delta(r) \rangle$ does not
depend on  $r$ due to the spatial uniformity in average, the use
of the sum rule  (2) gives
$$
\langle\Delta\rangle= g \langle\nu_F\rangle
\ln\frac{1.14\omega_0}{T} \langle\Delta\rangle \,,
\eqno(3)
$$
and the critical temperature $T_c$ is given by the BCS formula,
which contains the average density of states
$\langle\nu_F\rangle$. If the latter is maintained fixed, $T_c$
is not changed by disorder. However, self-averaging does not hold
in the general case, and deviations from the Anderson theorem
arise.

Equation (1) can be accurately solved for a small concentration of
the point-like impurities \cite{4}. This solution shows
possibility of two regimes. The first one
corresponds to the moderate variation of $\Delta(r)$ in space, so
that it remains more or less of the same order in the whole
space. The corresponding $T_c$ is given by the formula
$$
\frac{\delta T_c}{T_{c0}} = \frac{1}{\lambda L^d}\,
\int d^dr \, \frac{ \nu_0  \nu_1(r) +  \nu_1(r)^2}{\nu_0^2}
\,,
\eqno(4)
$$
where   $\nu_1({r})$ is a deviation of the local
density of states $\nu_F(r)$ from its unperturbed value $\nu_0$,
$\lambda=g\nu_0$ is the dimensionless coupling constant,
$T_{c0}$ is the transition temperature in the absence of disorder,
 $L^d$ is a volume for one impurity,
and integration is carried out over a vicinity of the single point
defect.  The linear in  $\nu_1(r)$ term exactly corresponds to the
Anderson theorem and relates the change in $T_c$ with the change
of the average density of states. Generally, $ \nu_1(r)$ is
comparable with $\nu_0$ and already Eq.4 predicts a possibility of
essential violation of the Anderson theorem. It is related with
the fact that the initially uniform order parameter is influenced
by point defects and can increase or decrease in their vicinity.

More essential deviations from the Anderson theorem arouse in the
second regime, when the order parameter is mainly localized at
the small number of "resonant" impurities producing the
quasi-local states near the Fermi level. The corresponding
estimate for $T_c$ \cite{4}
$$
T_c \sim g a^{-d} \sim \lambda E_0
\eqno(5)
$$
($a$ is the lattice spacing, $E_0$ is a scale of the order of the
Fermi energy or the bandwidth) is valid for small $g$ and
saturates by  a quantity of the order $\omega_0$, when $g$
increases.

Results (4) and (5), obtained for a small cocentration $c$ of
impurities, can be qualitatively extrapolated into the $c\sim 1$
region. It allows to give the adequate answer to the question in
the title of the paper:

(a) Disorder can enhance $T_c$ by a trivial reason, due to
increase of the average density of states.

(b) If the average density of states remains fixed, then $T_c$ can
be changed due to deviations from the Anderson theorem. This
effect is always positive in the framework of  formula (4).

(c) There is a possibility of the catastrophic increase of $T_c$
due to resonances on the quasi-local levels, though this regime
is affected by fluctuations (see below).

\vspace{2mm}

{\it Role of multifractality.} Recently there have been
claims \cite{5,6,7} that a great increase of $T_c$ is possible in
the vicinity of the Anderson transition due to multifractality of
wave functions. The analysis of \cite{5,6} is based on equation
$$
\Delta(\epsilon) = \frac{\lambda}{2}
\int\limits_{-\omega_0}^{\omega_0} \frac{I(\epsilon,\epsilon')
\Delta(\epsilon')} {\sqrt{\epsilon'^2+\Delta(\epsilon')}}
\tanh\left(\frac{\sqrt{\epsilon'^2+\Delta(\epsilon')}}{2T} \right)
\,d\epsilon' \eqno(6)
$$
with the kernel of the form
$$
I(\epsilon,\epsilon') =\left|\frac{E_0}{\epsilon-\epsilon'}
\right|^\gamma        \,,
\eqno(7)
$$
motivated by multifractal properties of wave functions. $T_c$ is
determined by equation (6) linearized in $\Delta$: accepting
$\Delta$ to be a function $\epsilon/T$ and dimensionalizing the
integral, one has for infinite $\omega_0$ \cite{5,6}
$$
T_c\sim E_0 \lambda^{1/\gamma} \,.
\eqno(8)
$$
The singular limit of small $\gamma$ was analyzed in \cite{1} and
lead to result
$$
T_c\sim E_0 \left[\left( \frac{E_0}{\omega_0}\right)^\gamma
+\frac{\gamma}{\lambda} \right]^{-1/\gamma}
\eqno(9)
$$
which reproduces (8) in the limit $\omega_0\to\infty$, the BCS
result in the limit $\gamma\to 0$ and describing saturation of (8)
by a value $\sim\omega_0$ for large $\lambda$. The claim of
\cite{1} that $\omega_0$ is always less than $E_0$ is incorrect:
$E_0$ can be small in semimetals and narrow band materials.

The linearized form of equation (6) is not equivalent to (1) (see
discussion in \cite{6}) and based on two assumptions: (i)
truncation of the Hamiltonian in the BCS spirit, and (ii)
averaging of the kernel independently of $\Delta$\,\footnote{\,It
means a self-averaging assumption, but in the modified form:
instead of the usual equation $\langle\Delta \rangle =\langle K
\rangle \langle\Delta \rangle$ one uses $\langle S\Delta \rangle
=\langle S K S^{-1}\rangle \langle S\Delta \rangle$  where $S$ is
a certain operator. It can be interpreted in the variational
spirit, considering $S$ as a kind of the trial function. }. Both
approximations are uncontrollable. The partial justification of
(i) was suggested in \cite{1}: truncation of the Hamiltonian is
rigorous for a pure superconductor and should be valid
approximately in the case of weak spatial inhomogeneity.  One can
partially excuse assumption (ii), using some effective exponent
in the capacity of $\gamma$.  Indeed, the kernel
$I(\epsilon,\epsilon')$ is determined by a matrix element $\int
d^dr |\psi(\epsilon, r)|^2 |\psi(\epsilon', r)|^2$ whose
estimation gives different values of $\gamma$ for the "usual" and
"typical" averaging \cite{6};
this
uncertainty is aggravated, if averaging is made with the weight
$\Delta(r)$.

The first argument explains why the strong localization regime
(corresponding to (5)) does not contain in (6) for
$I(\epsilon,\epsilon')=const$.  The second argument shows that a
difference between (5) and (8) arises on the level, which is not
controllable in the approach based on Eq.6. In fact, the
difference between (5) and (8) is not quantitative, but
qualitative. Result (5) is not restricted by a vicinity of the
mobility edge and remains approximately the same for any energy
inside the band; correspondingly, it has no relation to
multifractality. Nevertheless, the real physical mechanism is the
same for results (5) and (8) and related with resonances at
quasi-local levels \cite{4}. Indeed, if the local density of
states $\nu(\epsilon,r)$ is considered as a smooth function of
$\epsilon$, then its variation is finite: it corresponds to shifts
of the whole band by a value $W$ or $-W$ for the Anderson model
with distribution of site energies in the interval $(-W,W)$.
Unbounded fluctuations of $\nu(\epsilon,r)$, arising in the
context of multifractality, are necessarily related with a partial
discretization of the spectrum due to a presence of quasi-local
levels. If the usual value of $\gamma$ is exploited in (7), then
the $T_c$ value given by (5) is greater than (8); it means that
the Cooper instability occurs at configurations, which are
governed by individual peaks and not fractal clasters \cite{4}.

Weak multifractality considered in \cite{1} is practically actual
only for the $2D$ case in the regime of weak disorder. However,
this regime is described by formula (4), which can be obtained in
this case making two iterations of the Gor'kov equation and
exploiting the sum rule (2). In this case one can integrate along
all the system and use its whole volume in the capacity of $L^d$:
dividing of disorder into separate "impurities" is not
necessary. There is no need to use the approximate equation
(6), when the accurate result is available. By the
way, in the framework  of (4) the order parameter is proportional
to $\nu_F(r)$ and the logarithmically normal distribution for
$\Delta(r)$ \cite{1} follows trivially in the weak
multifractality regime.

It is clear from this consideration, that multifractality
is not a direct cause for the increase of $T_c$ and results of
kind (8) are related with the more universal mechanism.

\vspace{2mm}

{\it Role of fluctuations.} Equation (5) is a result of the
mean field theory. The corresponding configuration of the order
parameter is a uniform background $\Delta_0$ with abrupt peaks at
resonant impurities, whose concentration is of the order
$T_c/E_0$.  The order parameter can be considered as
positive\footnote{\,In the absence of magnetic effects,  the
kernel $K(r,r')$ is positive, and the Cooper instability
corresponds to  the nodeless eigenfunction.}, and so its phase is
the same in the whole volume.  When we come to a
fluctuational description, the modulus of the order parameter
remains practically unchanged, while the essential phase
fluctuations arise.  If the uniform background is neglected, then
the system is divided into practically independent
superconducting "drops", whose phases are fluctuating freely and
destroy the macroscopical coherence of the superconducting state.
If the uniform contribution $\Delta_0$ is taken into account,
the Josephson
coupling between drops arises and their phases become correlated.
The accurate fluctuational analysis of such a system is
nontrivial, but the general character of results is the same as
for the granular superconductors \cite{70}. If the ratio $T_c/E_0$
is not too small, then the resonant impurities are close to each
other and their Josephson interaction is strong enough for
stabilization of the mean-field solution at practically the same
$T_c$  value.  Contrary, if  $T_c/E_0$ is sufficiently small,
then the Josephson coupling between  drops is weak and
fluctuations destroy superconductivity at temperatures close to
the mean-field $T_c$ value. However,  decreasing of temperature
stimulates the growing of tails of the localized solutions
\cite{100}; the Josephson coupling between drops increases and
stabilizes superconductivity before $T_{c0}$  is reached. Hence,
fluctuations suppress $T_c$ in comparison with its mean-field
value, but do not eliminate enhancement of $T_c$ completely.

Analogous arguments can be given for configurations corresponding
to (8), where a fraction of the superconducting phase is estimated
as $(T_c/E_0)^\gamma$ \cite{6}. However, such configurations are
not actual, since the Cooper instablity corresponds to (5).

The paper \cite{1} suggests another way to deal with fluctuations.
Firstly, solution of (6) for $T=0$ is found, giving the spatially
inhomogeneous order parameter $\Delta_0(r)$. Secondly, the field
$T_c(r)$ of "local $T_c$" values is introduced, such as
$T_c(r)\propto \Delta_0(r)$.\,\footnote{\,In fact, this relation
is violated due to the presence of scale $T_{c0}$.}
 Finally, the global $T_c$ is defined
as a percolation threshold in the field $T_c(r)$. Such
percolation picture has a sense for certain conditions
\cite{101}, but it is not the case for weak spatial
inhomogeneity.

Indeed, the Gor'kov equation (1) defines the spatially
inhomogeneous configuration $\Delta(r)$, which appears at a
certain temperature.  This temperature, by definition, is a final
result for $T_c$:  there is no need to consider "local $T_c$" or
"percolation". Of course, one should attend for insignificance of
fluctuations, but this condition is rather weak. The Ginzburg
number is incredibly small for a pure superconductor; it increases
gradually with increasing of spatial inhomogeneity, but it is
possible to reach rather large values of ratio
$\Delta_{max}/\Delta_{min}$ before this number will approach unity
and fluctuations will become essential. Surely, no percolation is
necessary for weak spatial inhomogeneity. The percolation picture
becomes reasonable for the Ginzburg number of the order of unity,
when the mean-field estimate of $T_c$ is poorly defined and the
use of percolation allows to refine it.

\vspace{2mm}

{\it Role of interaction.} The BCS constant $\lambda$
corresponds to some effective interaction. In a more detailed
description it is combined from the electron-phonon coupling
and the Coulomb pseudopotential. The latter is known to increase
in the presence of disorder and it is the main cause for
$T_c$ degradation \cite{8,9}. This effect was not considered
explicitly in \cite{1,4,5,6}, but is surely essential in
discussion of the experimental situation. It is the main reason
why enhancement of $T_c$ is a rare thing in reality.

In paper \cite{7} the result analogous to (8) is obtained in the
framework of the Finkelstein renormalization group approach.
However, it is completely different from \cite{5,6} in the
initial assumptions and the discussed physical
mechanism. The Cooper constant $g$ is kept fixed in \cite{5,6},
but an attempt is made to advance beyond applicability of the
Anderson theorem.  Contrary, the authors of \cite{7} consider
renormalization of $g$ by disorder, while self-averaging is
accepted for granted. By the latter reason, the strongly
localized regime was not accessible in this approach, while the
obtained effect is lesser than (5). The questions also arise, how
results of \cite{7} agree with the usual picture of the Coulomb
pseudopotential enhancement.

\vspace{2mm}

In conclusion, disorder can enhance $T_c$: (a)  due to increase
of the average density of states; (b)  due to deviations
from the Anderson theorem; (c) due to resonances on the
quasi-local levels. The latter regime is affected by
fluctuations, and in some cases can be essentially suppressed.
Practically, enhancement of $T_c$ by disorder is a rare thing
due to the increase of the Coulomb pseudopotential.
\vspace{2mm}

The author is indebted to M.V.Sadovskii for the discussion of
paper \cite{1}.


\vspace{10mm}

\begin{center}
{\large\bf Reply to arXiv:1502.06282}

\vspace{3mm}

\parbox{70mm}{\footnotesize \,As a reaction to the preceding
overview \cite{13}, Mayoh and Garcia-Garcia have submitted the
comment \cite{14}. This comment has a form of the personal attack
and contains the whole series of untrue statements. We give brief
remarks, in order to reveal these statements.  } \end{center}

\vspace{5mm}

1. Our mean field results for a superconductor with a small
concentration of the point-like impurities \cite{4}
are based on the accurate solution of the Gor'kov
equation.  These results still persists, even if somebody
does not like them.  Contrary, the approach of \cite{1}
involves uncontrollable approximations \cite{13}.

\vspace{2mm}

2. It is repeatedly stated in \cite{14} that we predict $T_c\sim
5000 K$ for the superconducting transition temperature, which "has
never, and very likely will never, been found numerically or
experimentally".  In fact, it was clearly indicated (see the text
after Eq.2 in \cite{4}) that $T_c$ is bounded by the quantity
$\omega_0/\pi$ ($\omega_0$ is the cut-off frequency), which for
the phonon mechanism corresponds to values already observed for
oxide superconductors ($T_c \sim 150 K$, $\omega_0\sim 400 K$).  A
possibility to describe the latter in terms of our model was
discussed in Sec.7 of \cite{4}.

\vspace{2mm}

3. The authors of  \cite{14} write on "unjustified use of the mean
field approximation". In fact, insufficiency of the mean field
approach was indicated in \cite{4,13} and the role of fluctuations
was extensively discussed.  We cannot help, if our arguments are
ignored.

\vspace{2mm}

4. According to \cite{14}, our formula (4) in \cite{13}
is related with
"poorly defined" quantities and unclear conditions of
applicability. In fact, all explanations were given in
\cite{4,13} and we can repeat those of them that are questioned
in \cite{14}:  $L^d$ is a volume for one impurity, i.e. the whole
volume divided for a number of impurities; the quantity
$\nu_1(r)$ is the difference $\nu_F(r)-\nu_0$, where $\nu_F(r)$
is the local density of states, defined in a standard manner (see
Eq.8 in \cite{4}), and $\nu_0$ is its unperturbed value. For a
small amplitude of disorder ($|\nu_1(r)|\ll \nu_0$) the indicated
formula can be obtained by two iterations of the Gor'kov equation
\cite{13}.  It remains valid for a small concentration of strong
impurities, when $|\nu_1(r)|\sim \nu_0$  \cite{4}. It is not
restricted by any assumption on the correlation between
impurities.

\vspace{2mm}

5. According to \cite{14}, "Suslov claims $\ldots$ that we do not
discuss phase-fluctuations".  There is no  such statement in
\cite{13}; our comments on the role of interaction are brief and
contain no criticism of \cite{1}.

\vspace{2mm}

6. According to \cite{14}, there is a wrong statement in
\cite{13}: "Suslov also states that we claim $E_0$ is always
larger than $\epsilon_D$". One can compare it with the text in
 \cite{1}:  "We believe that this is
necessary since $\epsilon_D\ll E_0$ so it is inconsistent to take
the Debye energy to infinity while keeping $E_0$ finite". This
argument was used in \cite{1} to reject the result $T_c\sim E_0
\lambda^{1/\gamma}$ appearing in preceding publications: "This is
an expression that, we are at pains to stress, is not recovered
in our formalism" \cite{14}. In fact, this result follows from
Eq.22 of \cite{1} in the limit $\epsilon_D\to \infty$,
independently of the desire of authors.

\vspace{2mm}

7. According to \cite{14}, "Suslov claims that our results are
only valid in small and strictly two dimensional systems", while
they are valid also for thin films and two-dimensional systems
with spin-orbit interactions.  In fact, it is written in \cite{13}
that "weak multifractality considered in [1] is practically
actual only for the $2D$ case in the regime of weak disorder",
and there is no rejection of the indicated additional
applications.

\vspace{2mm}

8. Our main criticism of \cite{1} refers to the use of the
percolation picture \cite{101} beyond the limits of its
applicability. Indeed, the case of the "moderate spatial
inhomogeneity" is described by the Gor'kov equation, which has
solution $\Delta(r)$ arising at a certain temperature.  This
temperature, by definition, is a final result for $T_c$:  there is
no need to consider "local $T_c$" or "percolation". Such situation
persists, till fluctuations are insignificant and the mean field
estimation of $T_c$ is well-defined.  The percolation picture
becomes reasonable for "very strong spatial inhomogeneity", when
the Ginzburg number becomes of the order of unity; in this case
the mean-field estimate of $T_c$ is poorly defined and the use of
percolation allows to improve it.

The authors of \cite{1} try to disprove our arguments basing
on the difference between "weak inhomogeneity" and "weak
multifractality". This attempt is not successful due to the
facts:

(a) The suggested in \cite{1} partial justification of Eq.6 in
\cite{13}  refers to "weak inhomogeneity" and not to "weak
multifractality".

(b) It is clear from \cite{13} that for applicability of
percolation, the ratio $\Delta_{max}/\Delta_{min}$ should be
greater than some large parameter; here $\Delta_{max}$ and
$\Delta_{min}$ are typical (not exclusive) values. The
distribution for $\Delta(r)$ is presented in Fig.4 of \cite{1}.
This distribution can be cut-off on both sides, since its tails
correspond to local perturbations, which have no consequences
for global superconductivity. After that the ratio
$\Delta_{max}/\Delta_{min}$ is typically of the order of unity: it
corresponds to "moderate inhomogeneity" and applicability of the
Gor'kov equation.

(c) The situation can be discussed constructively. Practically,
"weak multifractality" corresponds to the weakly disordered 2D
case, which is described by formula (4) in \cite{13}.  The
corresponding order parameter is either slightly perturbed (for
weak impurities), or its perturbations are local (for a small
concentration of strong impurities). In both cases, there are no
problems with fluctuations, and hence there is no place for
percolation.

The authors of \cite{14} write correctly that "according to Suslov
our percolation analysis is intended to describe fluctuations": it
is right, in spite of their objections. They are wrong in
ascribing to us an idea that "the value of $T_c$ resulting from
percolation is similar to that obtained by averaging over
$T_c(r)$".

\vspace{2mm}

 9. It is stated in \cite{14} that "in weakly coupled
superconductor $\lambda\ll 1$ so $T_c/E_0$ is always small and the
global critical temperature must be necessary zero". This
statement reveals a complete misunderstanding of our arguments. We
say that, for not very small ratio $T_c/E_0$, the average distance
$a\left(E_0/T_c\right)^{1/3}$ between resonant impurities becomes
comparable with $a$ and the background value $\Delta_0$ of the
order parameter becomes comparable with its resonant peaks. Then
our "strongly localized regime" becomes "moderately localized" and
has no problems with fluctuations. In any case, we do not restrict
ourselves by weak coupling,
and in the worst situation $T_c$ falls to $T_{c0}$ and not to
zero.

\end{document}